# Investigating the CVD synthesis of graphene on Ge(100): towards layer by layer growth

A.M. Scaparro,[†] V. Miseikis,[‡] C. Coletti,[‡] A. Notargiacomo,[§] M. Pea,[§] M. De Seta,[*,†,⊥] and L. Di Gaspare[†,⊥]

[†]Dipartimento di Scienze, Università degli Studi Roma Tre, Viale Marconi 446, 00146 Rome, Italy

[‡]Center for Nanotechnology Innovation @NEST, IIT, Piazza San Silvestro 12, 56127 Pisa, Italy

[§]Institute for Photonics and Nanotechnology, Via Cineto Romano 42, 00156, CNR-Rome, Italy

ABSTRACT

Germanium is emerging as the substrate of choice for the growth of graphene in CMOS-compatible processes. For future application in next generation devices the accurate control over the properties of high-quality graphene synthesized on Ge surfaces, such as number of layers and domain size, is of paramount importance. Here we investigate the role of the process gas flows on the CVD growth of graphene on Ge(100). The quality and morphology of the deposited material is assessed by using $\mu$-Raman spectroscopy, x-ray photoemission spectroscopy, scanning electron and atomic force microscopies. We find that by simply varying the carbon precursor flow different growth regimes – yielding to graphene nanoribbons, graphene monolayer and graphene multilayer – are established. We identify the growth conditions yielding to a layer-by-layer growth regime and report on the achievement of homogeneous monolayer graphene with an average intensity ratio of 2D and G bands in the Raman map larger than 3.

INTRODUCTION

Since its isolation, graphene has attracted a huge interest due to its unique electronic structure and exceptional electronic and mechanical properties. Scalable and controlled synthesis of high quality graphene is the main request for several potential applications. Chemical Vapor Deposition (CVD) has been predicted as the leading growth technique for large-scale deposition of graphene films[1] and has emerged as the dominant method to synthesize continuous films of "electronic grade". Even though CVD graphene on metal has reached high standards of quality,[2-6] the compatibility with mainstream silicon microelectronics requires the growth of metal-free graphene on CMOS compatible semiconductor[1,7]. Germanium is a perfect candidate as substrate since high quality single crystal epitaxial Ge on Si is routinely available[8] and it does not form a stable carbide[9]. The successful growth of graphene on Ge could enable the integration in the CMOS platform of graphene-based optoelectronic devices as high performance transistors,[10-12] terahertz emitters,[13] and electro-optical modulators[14].

Large area single crystal graphene films on the Ge(110) substrate are achieved[15] exploiting the anisotropic surface symmetry of Ge(110)[16]. However, the development of a CMOS integrated graphene electronics requires the successful growth of high quality graphene on the Ge(100) surface. In Refs.[17-19] the CVD growth of graphene on Ge(100) substrates and Ge/Si(100) was demonstrated. Despite the variety of graphene features reported in literature,[17-22] the process conditions yielding to the development of monolayer, multilayer, or graphene nanoribbons have not been yet well established to-date, hindering the development of wafer scale graphene-based optoelectronics, where the control over the morphology and structure of the deposited graphene is mandatory[23,24].

The catalyzed CVD growth of graphene is determined by a specific sequence of processes, i.e. the formation of active carbon species on the substrate, the graphene domain nucleation, and the domain lateral growth. In all these processes hydrogen plays an active role, promoting the activation of surface-bound carbon and acting as an etching agent that controls the size and morphology of the resulting graphene domains.[25,26] The different relevance of these processes at different stages of the growth depends essentially on the ratio in the process gas mixture of the $H_2$ carrier gas with the carbon-containing hydride (e.g. $CH_4$) on the substrate surface, where the catalysis and diffusion of the C species take place.

In this work we investigate the processes at the basis of the different CVD growth regimes of graphene on Ge(100) by varying the methane flow, the $H_2$:$CH_4$ flow ratio, and the deposition time. The joint use of complementary experimental techniques, both spectroscopic and morphological, allowed an exhaustive characterization of the "as grown" materials (without the need of graphene transfer on other substrates) at different stages and conditions of the synthesis. We demonstrate that by simply varying the methane flow is possible to tune the growth so to obtain either graphene nanoribbons, graphene monolayer and graphene multilayer. The role of the different processes bringing to the formation of graphene on Ge(100), from nucleation to domain lateral growth and etching, domain merging and multilayer development in the different growth conditions are established. Notably, this work leads to the identification of a layer-by-layer growth regime which allows for the synthesis of high quality monolayer graphene in a controllable and reliable manner. As the presented process is developed in a commercially available reactor, it might be easily reproduced and its adoption in industrial production appears feasible. The report of a layer-by-layer growth on Ge substrates and in a commercially available reactor is a first step towards a larger accessibility of high quality graphene whose properties can be tuned by controlling the number of layers.

EXPERIMENTAL METHODS

Graphene was grown on Ge(100) substrates by employing a commercially available 4" cold-wall CVD reactor (Aixtron BM). The Ge(100) substrates were cut into 1x1 $cm^2$ pieces from a 4" wafers (N-type Sb-doped, $n=10^{16}$ cm$^{-3}$, <100>+/-0.1°, double side polished) and cleaned with ultrasonic bath in isopropyl alcohol followed by rinsing in de-ionized water. The substrates were then placed at the center of the BM growth chamber and heated up to the deposition temperature in $H_2$ and Ar gas mixture. Once the growth temperature was reached, the carbon precursor (i.e., $CH_4$) was introduced in the chamber. In the reported experiments $H_2$ and Ar flows were set at 200 and 800 sccm respectively, while the $CH_4$ flow varied in the 1-10 sccm range. The growth was performed at 100 mbar and at T=930°C, only few degrees below the Ge surface melting. As a matter of fact good quality graphene on Ge can be achieved only employing growth temperatures in a small range close to the Ge melting. To ensure good reproducibility and a homogeneous temperature on the whole Ge surface the substrate heating is carried out by a multi-step temperature ramp. After graphene deposition the system was cooled down to room temperature in $H_2$ and Ar. No differences were appreciated performing the cooling process in either an $H_2$:Ar mixture or in pure Ar.

The samples were characterized by Raman spectroscopy, scanning electron microscopy (SEM), atomic force microscopy (AFM), and x-ray photoelectron spectroscopy (XPS).

Raman spectroscopy (Renishaw inVia confocal Raman microscope) was performed using an excitation wavelength of 532 nm, a 100x objective, a laser spot size of 1 $\mu$m and a step size of 1 $\mu$m for area mapping. The intensity ratio of the 2D, G and D bands were evaluated by using the integrated area of the peaks.

The sample morphology was investigated by means of SEM (FEI Helios 600 Nanolab DualBeam) and AFM (Veeco CP-II) operating in tapping mode. The $x_n$ coverage fraction of domains with n-layers of graphene (n-LG) was evaluated from a standard analysis of the SEM images performed by setting user-defined threshold levels for the grey scale intensity. For all the samples n-LG domain size distribution, domain average size (defined as the square root of the area) and its standard deviation were determined. The domain size analysis has been performed also on AFM data on the samples that do not exhibit the Ge nano-texturing induced by the graphene growth. In this case the threshold intervals for the height values associated to n-LG were extracted from line-profile analysis. The results of the two methods are in good agreement in all the analyzed samples.

The XPS measurements were carried out using a monochromatic Al Kα source ($h\nu$=1486.6 eV) and a concentric hemispherical analyzer operating in retarding mode (Physical Electronics Instruments PHI), with overall resolution of 0.4 eV. The C1s core level area intensity of each sample has been normalized to that acquired in the same experimental conditions on a commercial single layer graphene (SLG), a CVD-deposited graphene on copper foil and transferred on a $SiO_2$ substrate, purchased from Graphenea.

RESULTS

**$CH_4$ flow influence on the growth kinetics**

To gain insight into the graphene CVD growth on the Ge(100) surface we investigated the influence of the $CH_4$ flow on the growth kinetics by synthesizing graphene films at fixed deposition time $t_D$=60 min varying the $CH_4$ flow F in the 1-10 sccm range. The resulting $H_2$:$CH_4$ flow ratio $R$ varied from $R$=200 down to $R$=20.

The Raman spectra of the samples grown at different F and their analysis are shown in Figure 1. In all the samples the Raman spectra exhibit the typical graphene features, i.e. 2D and G bands located at ~ 2700 and ~ 1600 $cm^{-1}$, and the D peak related to the presence of residual defects in the film at ~1350 $cm^{-1}$ (see Figure 1a).

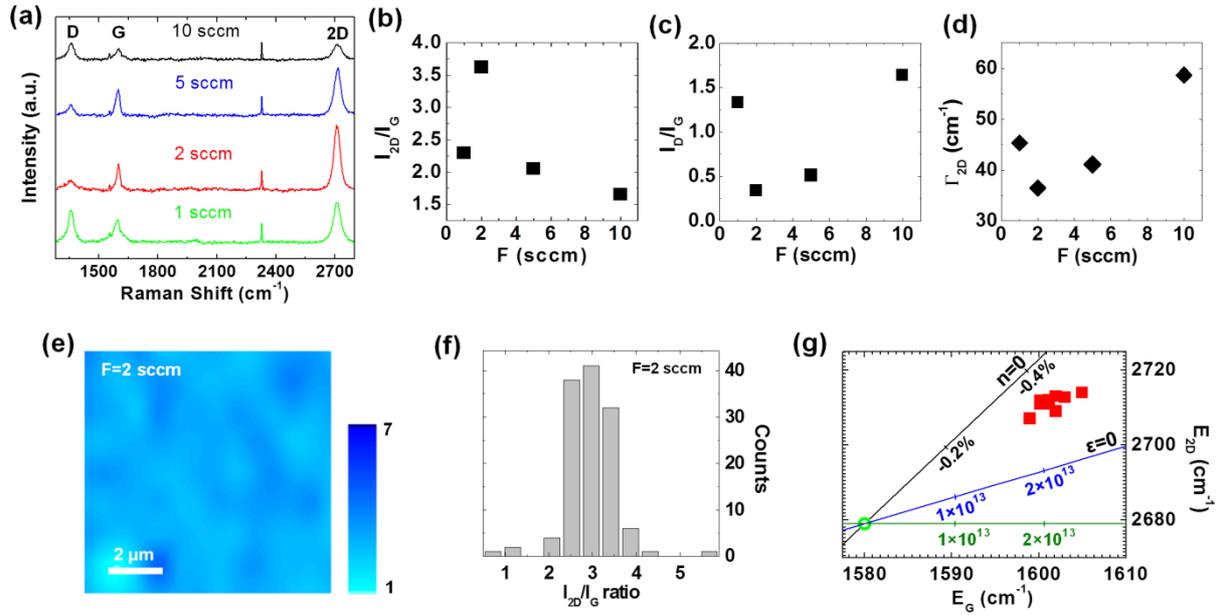

**Figure 1** (a) Raman spectra of samples grown at different $CH_4$ flows F for $t_D$=60 min; (b) $I_{2D}/I_G$, (c) $I_D/I_G$ intensity ratios and (d) $\Gamma_{2D}$ as a function of the methane flow F; (e) $\mu$-Raman map and (f) related histogram of $I_{2D}/I_G$ intensity ratio of the single layer graphene grown at F=2 sccm and $t_D$=60 min; **(g)** Plot of the 2D vs G-band for the graphene film deposited at F=2 sccm. Colored lines indicate $E_{2D}$ and $E_G$ relationship for strained undoped (n=0, black line), unstrained p-doped ($\varepsilon$=0 blue line) and unstrained n-doped ($\varepsilon$=0 green line) graphene. The neutrality point (circled green point) corresponds to the expected 2D and G positions for suspended freestanding single-layer graphene.

The SEM and AFM images of the samples displayed in Figures 2 and 3 point out that the variation of F in this range gives rise to completely different growth regimes.

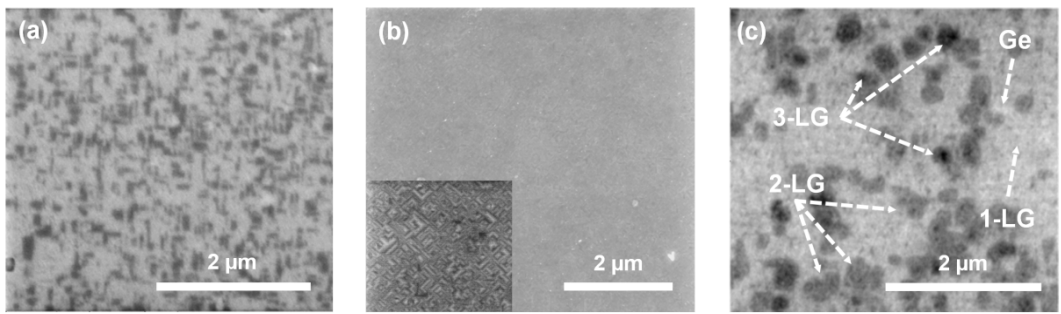

**Figure 2** SEM images of samples grown at different $CH_4$ flows F for $t_D$=60 min. (a) F= 1 sccm; (b) F=2 sccm ; (c) F=5 sccm. Inset in panel (b) reports the SEM image of the sample at F=2 sccm acquired at higher magnification. The panel and its inset have the same scale bar. In panel (c) the arrows mark regions having different number of graphene layers that appear with different grey scale intensity.

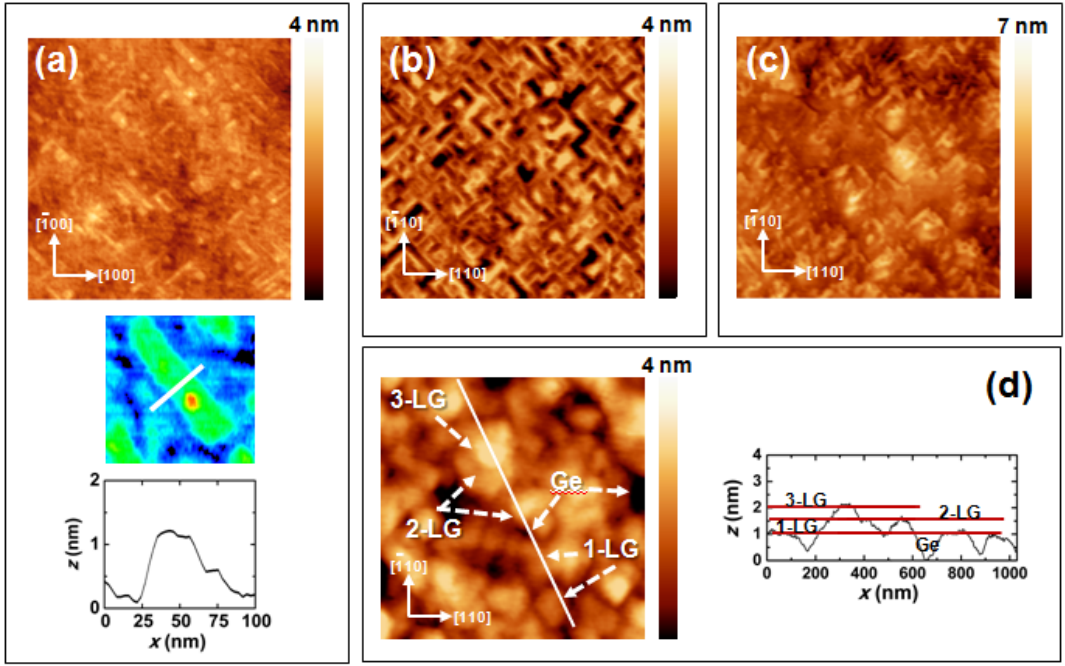

**Figure 3** AFM topographies of samples grown at different CH$_4$ flows F for t$_D$=60 min. All images have 1 μm scan size. (a) F= 1 sccm. The detail (image size 160×160 nm$^2$) of a single nanoribbon and its line profile are also reported. (b) F=2 sccm . (c) F=5 sccm. (d) F=10 sccm. The arrows mark regions having different number of graphene layers. The height profile along the white line in the AFM topography is also shown. In panel (a) the AFM scan is along the [100] direction to better detect the nanoribbons oriented along the <110> directions.

For F=1 sccm (Figures 2a and 3a) graphene nanoribbons oriented along the <110> directions and mutually perpendicular develop on the surface in a way similar to that reported in Ref. [20]. Their asymmetric shape characterized by a relatively high length/width aspect ratio is promoted by the Ge(100) surface anisotropy.[21] Nanoribbons cover about 35 % of the Ge surface and have an average aspect ratio α=6 and an average length R$_l$=150 nm. Their typical shape measured by AFM is highlighted at the bottom of Figure 3a. Its height of 1 nm is consistent with a single layer feature.[17] The relatively high intensity ratio between D and G peaks (I$_D$/I$_G$) (see Figure 1c) in the Raman spectrum is ascribable to the small size of nanoribbons. As a matter of fact, the nanoribbon boundaries give a large contribution to the defect-related D peak being their size smaller with respect to the laser spot diameter (~1μm). If the deposition time t$_D$ is doubled keeping the same growth conditions, the surface coverage increases to about 60%. In some regions of the sample the nanoribbons merge and the $I_D/I_G$ intensity ratio decreases down to ~0.7.

For F=2 sccm a single layer graphene film covers completely the Ge substrate. As a matter of fact, the SEM image of Figure 2b shows a uniform graphene surface without domains. A nano-texture of the Ge surface underneath the graphene monolayer is visible in the SEM image acquired at large magnifications reported in the inset of Figure 2b and in the corresponding AFM topography reported in Figure 3b. We can observe the presence of hill-and-valley faceted structures oriented along the two mutually perpendicular <010> directions. The average peak-to-valley height is 3 nm and the facet slope ~5°, compatible with a {0,1,10} facet. It is worth noticing that in a simulated growth (i.e. exposing the Ge substrate for $t_D$=60 min only to $H_2$ and Ar with F=0) the Ge substrate shows a flat surface (rms roughness ~0.3 nm) with the typical Ge terraces having height of about 0.14 nm. A similar nano-faceting of the Ge surface underneath a continuous graphene film is reported in Refs[19,20]. Interestingly, in our growth conditions the Ge nano-faceting appears only in samples where a continuous or a quasi-continuous graphene film is present. A possible explanation is that the nanofaceting is due to the development of local strain of the Ge surface induced by the growth of large enough and ordered graphene domains. The single layer nature of the graphene film grown at F=2 sccm is confirmed by the Raman analysis shown in Figure 1. The Raman 2D-peak of this sample has a narrow single Lorentzian line shape with a FWHM $\Gamma_{2D}$ of 36 cm$^{-1}$. The residual integrated $I_D/I_G$ intensity ratio ~0.3 can be related to the polycrystalline nature of the graphene grown on Ge(100) or to the presence of isolated defects.[27] The $I_{2D}/I_G$ intensity ratio $\mu$-Raman map and the relative histogram are reported in Figure 1e and 1f, respectively. The $I_{2D}/I_G$ distribution has an average value and a standard deviation equal to 3.1 and 0.55, respectively. It is well known that the $I_{2D}/I_G$ ratio decreases with the graphene charge doping.[28] Raman spectroscopy can be used to evaluate the doping level and the amount of strain in graphene by analyzing the 2D- and G-band positions.[29] In order to estimate the charge density in Fig. 1g we report the 2D vs G peak positions in Raman spectra taken in different surface regions of the sample grown at F= 2 sccm. The charge density can be estimated using the strain and doping deconvolution reported in Ref. [29]. The figure shows that the experimental data are not positioned on the charge neutrality line. Although the analysis does not allow to distinguish between holes and electrons we can estimate a charge density of the order of $1 \times 10^{13}$ cm$^{-2}$. This value is comparable with the electron density found in Ref. [19] by Hall measurements. The same analysis reveals also the presence of a compressive biaxial strain $\varepsilon \sim -0.3\%$ similar to that observed in Ref. [19]. We point here out that in our samples the obtained average $I_{2D}/I_G$ intensity ratio is larger and the standard deviation is smaller than those reported in literature[17-19] demonstrating the high quality and uniformity of our single layer graphene.

For F=5 sccm single, bi- and trilayer graphene domains coexist on the Ge surface as shown in Figure 2c. The growth mode observed in this condition is similar to that reported in Ref.[18]. The average bilayer and trilayer domain sizes are 330±60 nm and 170±70 nm, respectively. Less than 10% of the Ge surface is left exposed. The measured Raman features ($I_{2D}/I_G$ =2 and $\Gamma_{2D}$= 41 cm$^{-1}$) are compatible with the presence of mono- and multilayer graphene domains. The AFM image of Figure 3c shows that the Ge substrate is nano-textured. The height oscillations of the nano-faceting are larger than the graphene layer spacing and do not allow a reliable detection of n-LG domains from the AFM topography.

A further increase of the CH$_4$ flow results in a poorer quality of the graphene films. For F=10 sccm, the Ge nanofaceting is not present. Domains having a smaller average lateral size of ~100 nm, irregular shapes, and different heights (0-, 1-, 2-, 3- LG) are observed (see the AFM topography and the relative height profile in Figure 3d). The Raman analysis evidences the poor quality of the deposited material in this condition as demonstrated by the high values of $I_D/I_G = 1.6$ and $\Gamma_{2D} = 59$ cm$^{-1}$.

The morphology variation induced by the use of different CH$_4$ flows is quantified in Figure 4a, where the coverage fractions $x_n$ of 0- (uncovered Ge), 1-, 2- and 3-LG regions, as determined by the analysis of SEM and AFM images, are reported as a function of F. The $x_n$ values are further validated by the analysis of the XPS data. As a matter of fact the integrated intensity ($I_C$) of the C1s peak of each sample normalized to that measured on a commercial SLG ($I_C^{SLG}$) is related to the coverage fractions $x_n$ by the expression:

$$\frac{I_C}{I_C^{SLG}} = \sum_{n=1}^{n_{max}} x_n \frac{1-e^{-\frac{n}{\lambda_{eff}}}}{1-e^{-\frac{1}{\lambda_{eff}}}} \qquad (1)$$

where $n$ is the number of the graphene layers, $\lambda_{eff} = \lambda \cos\vartheta = 3.2$ monolayers[30] is the effective electron escape depth, $\vartheta = 30°$ is the detection angle, i.e. the angular displacement of the detector slit measured respect to the normal to the sample surface.

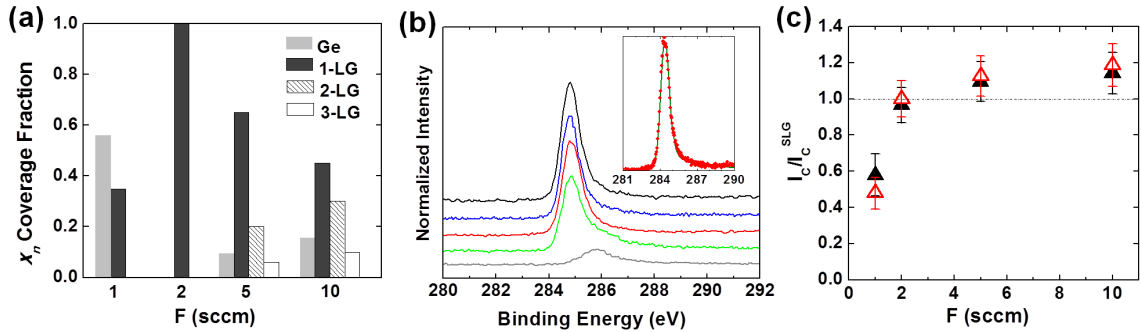

**Figure 4** (a) Coverage fraction $x_n$ of uncovered Ge surface, 1-, 2- and 3-LG present in the samples grown at different F for $t_D = 60$ min. (b) High resolution C1s core level spectra: from top to bottom: F=10, 5, 2, 1 sccm. The spectra have been normalized to the emission intensity of a standard commercial SLG. The bottom grey spectrum is taken on the Ge substrate exposed only to H$_2$ and Ar without methane in the growth chamber (F=0): it has a symmetric line shape and is centered at a higher binding energy (~285.8 eV), corresponding to sp$^3$-bonded carbon due to contamination. In the inset, the fit (continuous green line) performed on the C1s peak of sample grown at F=2 sccm (red dots) is reported. In (c) the experimental $I_C/I_C^{SLG}$ intensity ratio (full black triangles) are compared with the values (open red triangles) calculated by eq. (1) using the $x_n$ coverage fraction reported in panel (a). For the sample with F=1 sccm the C1s component due to the C contamination is not included in $I_c$.

The high resolution C1s XPS spectra and the experimental $I_c/I_c^{SLG}$ ratios are displayed in Figure 4b and 4c, respectively. The spectra of samples deposited at F=2, 5, 10 sccm exhibit the typical C1s graphene asymmetric line shape[31] peaked at 284.4 eV and well fitted with a Doniach-Sunjic profile with an asymmetry parameter of 0.12. The fit performed on the sample grown at F=2 sccm is reported in the inset of Figure 4b. Its overall C1s emission area agrees within ~5% with that measured in the same experimental condition on a commercial SLG, thus confirming the single layered nature of the sample grown at F=2 sccm. The absence of a C1s component at lower energy indicates that Ge-C bonds are negligible. Therefore graphene interacts with the Ge substrate through Van der Waals forces. For F=1 sccm, i.e. when nanoribbons partially cover the Ge surface, an additional carbon component at higher binding energy $E_b$= 285.8 eV due to C contamination appears. As a matter of fact, a peak with the same line shape and binding energy is present in the XPS spectrum of a Ge substrate exposed for $t_D$=60 min only to hydrogen and argon, i.e. at F=0. This C contamination component appears in the XPS spectra whenever a significant fraction of the germanium surface remains uncovered.

In Figure 4c we compare the experimental $I_c/I_c^{SLG}$ ratios to the theoretical ones calculated using the $x_n$ values of Figure 4a and eq.(1). Their good agreement for each sample confirms the reliability of the $x_n$ values estimated from morphological data.

**Growth time evolution of graphene layers at CH$_4$ flow F=2 sccm**

The above reported analysis demonstrates that the use of F=2 sccm and $R$=100 corresponds to a "special" growth condition in which the combined effect of carbon activation, graphene domains growth, and etching is optimized for the development of a continuous graphene monolayer. Therefore, we have selected these process conditions to investigate the growth evolution as a function of the deposition time.

The morphology of the films deposited for $t_D$=30 and 120 min, i.e. before and after the development of the continuous graphene monolayer discussed in the previous section, is shown in Figure 5.

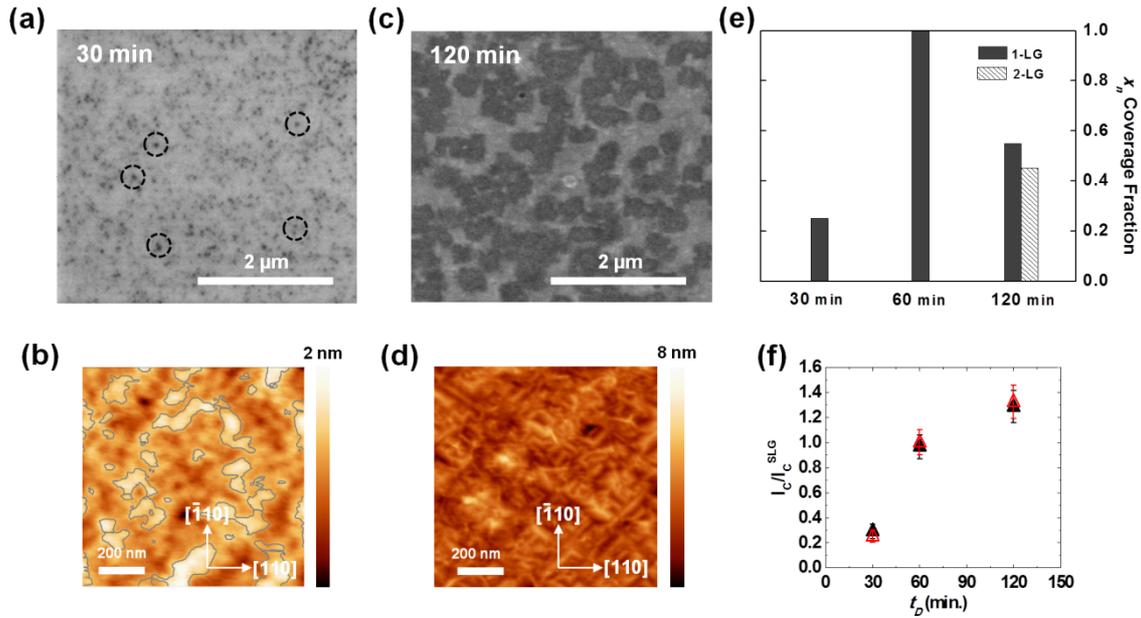

**Figure 5** (a) SEM image and (b) AFM topography of the sample grown for $t_D$=30 min at F= 2sccm. The black circles in panel (a) highlight the small graphene grains present in the early stage of the growth that cover ~25% of the surface. In panel (b) the grey lines mark regions having height ~ 1 nm. (c) SEM image and (d) AFM topography of the sample grown for $t_D$=120 min at F= 2sccm. (e) Coverage fraction $x_n$ of 1- and 2-LG of samples grown at F=2 sccm as a function of the deposition time. In (f) the experimental $I_C/I_C^{SLG}$ intensity ratio (full black triangles) are compared with the values (open red triangles) calculated by eq. (1) using the experimental $x_n$ coverage fraction reported in panel (e).

For $t_D$=30 min the SEM image (Figure 5a) evidences the presence of small graphene grains on the Ge surface with a coverage fraction of about 25%. They have an average size of 55±20 nm and a density of ~10÷20 $\mu m^{-2}$. The AFM topography (Figure 5b) demonstrates that at this stage of the growth the Ge surface is not nanostructured and has a rms surface roughness of 0.26 nm. The surface area where graphene grain thickness is ~1 nm (delimited by the grey lines in the AFM image) is consistent with the 25% coverage of graphene grains evidenced by SEM.

The Raman spectrum of the sample (Figure 6) reflects with the "seed-like" stage of the graphene domain evolution. As a matter of fact, the defect band D has very high intensity ($I_D/I_G=1.9$). Although the $I_{2D}/I_G=3$ confirms the monolayer nature of the grains the width of the 2D bands is relatively large ($\Gamma_{2D}=51.8$ cm$^{-1}$) suggesting that defects impact also this parameter.

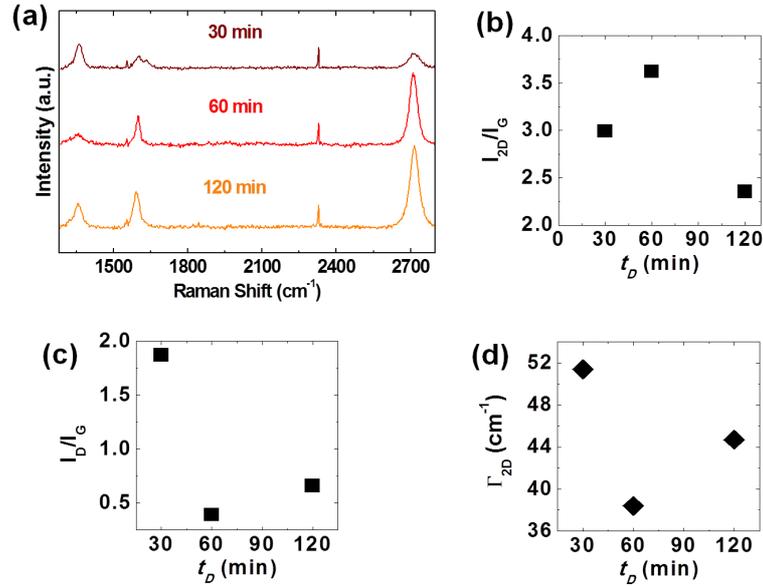

**Figure 6** (a) Raman spectra of samples grown at F=2 sccm for different deposition time. (b) $I_{2D}/I_G$, (c) $I_D/I_G$ intensity ratio and (d) $\Gamma_{2D}$ as a function of the deposition time.

For $t_D=60$ min these graphene grains evolve towards the continuous graphene monolayer, whose characteristics have been described in the previous section. A further increase of the deposition time brings to the development of 2-LG domains. The SEM image reported in Figure 5c shows that for $t_D=120$ min 2-LG domains cover 45% of the surface, suggesting that the average growth rate of the second layer is lower than that of the first one and/or a certain rest period is spent between the completion of the first layer and the nucleation of the second one. The average size of 2-LG domains is 350±50 nm. The presence of 3-LG domains is negligible although the equivalent thickness of this sample is larger with respect to that of the sample grown 60 min at F=5 sccm, where multilayer domains were found. Also in this case the Ge surface is nanofaceted, as evident in the AFM topography shown in Figure 5d. The presence of 2-LG domains impacts on the Raman feature: $I_{2D}/I_G$ decreases below 3 while $I_D/I_G$ and $\Gamma_{2D}$ becomes larger than the corresponding values of the graphene monolayer grown at $t_D$=60min.

In Figure 5e the time evolution of the coverage fractions $x_n$ of 1- and 2-LG regions determined by the morphological analysis is displayed. Once more, these data are validated by XPS analysis: the experimental $I_C/I_C^{SLG}$ intensity ratios and the corresponding theoretical values (eq.(1)) evaluated by taking into account the measured coverage fraction $x_n$, agree within experimental errors at all growth time (Figure 5f). This time evolution points out that at F=2 sccm the growth proceeds in a layer by layer regime, with formation of continuous single layer graphene by domain coalescence and the development of the subsequent one starting after the complete coalescence has occurred. The quantitative Raman analysis (Figure 6b-d) confirms the growth evolution evidenced by morphological and XPS data, showing a minimum in both $I_D/I_G$ and $\Gamma_{2D}$ and a maximum of $I_{2D}/I_G$ at $t_D$=60 min. The achievement of a layer-by-layer growth regime could represent a breakthrough toward the deposition of high quality graphene with controllable number of layers on CMOS compatible semiconductor substrate.

DISCUSSION

In the CVD growth of graphene on metallic surfaces the catalysis of methane on the surface and the successive formation of $CH_x$ active species enhanced by the presence of H produce a carbon-adatom species supersaturated surface, where the graphene nucleation takes place.[32] Graphene nuclei grow further by consuming the adsorbed carbon species. The occurrence of successive nucleation depends on the balance between the production rate of carbon growth species from catalyzed methane decomposition and their consumption rate due to nucleation and growth or to recombination and desorption. The domain lateral growth rate and shape are also influenced by the H etching that removes the C atoms bound to the graphene boundaries.

Our findings suggest that the same growth model reported in Ref. [32] can be applied to the CVD growth of graphene on Ge. We attribute this to the similarities existing between the C-Ge and C-Cu material systems: similar catalytic activity, extremely low solubility of carbon, absence of a stable carbide. [9,17]

The time evolution of the $x_n$ coverage fractions and of the Raman spectra points out that at F= 2 sccm and $R$=100 the growth proceeds in a layer-by-layer regime suggesting that in these process conditions the C adatom species ($CH_x$) concentration before nucleation is just above the critical supersaturation level. The nucleation and grow th of graphene grain deplete the adsorbed carbon species and their concentration is reduced to a level where the nucleation rate can be negligible and only monolayered domain enlargement takes place up to coalescence. After domain merging and completion of the first graphene layer the initial supersaturation condition leading to novel nucleation is restored and the second graphene layer begins to form. The high reproducibility of the continuous graphene monolayer achievement at $t_D$=60 min in several deposition runs points to the occurrence of a rest period between the completion of the first layer and the nucleation of the second one, that is to restore the supersaturation conditions. At present, our data do not allow to unambiguously conclude whether the 2-LG nucleates on top (wedding cake structures WC) or underneath (inverted wedding cake structures IWC) the first one.

For low $CH_4$ flow (F=1 sccm) the graphene growth slows down. As a matter of fact, the growth rate at F=1 sccm ($5.4 \times 10^{-3}$-LG/min) is significantly smaller than half the rate at F=2 sccm ($1.7 \times 10^{-2}$-LG/min). Taking into account that the higher $H_2$:$CH_4$ ratio should increase the production rate of carbon growth species[25] the sub-linear behavior of the growth rate can be ascribed to the effect of hydrogen etching. Although hydrogen etching has the capability of "drawing" the shape of the graphene domains[25], the role of hydrogen on the development of the graphene nanoribbons observed at F=1 sccm is still not clear. As a matter of fact graphene nanoribbons were recently obtained in UHV by using ethylene without H$_2$.[21] Comparing the previously published data[20,21] with our findings, we can observe that all these nanoribbon growth regimes are characterized by a small growth rate.

For the highest methane flows investigated (F≥5 sccm), we found the coexistence of uncovered, 1-, 2- and 3-LG regions on the Ge surface. In these growth conditions the increase of C adatom species concentration leads to a strong supersaturation on the Ge(100) surface. The production rate of carbon growth species by catalyzed methane decomposition is much larger than their total consumption rate allowing the further nucleation of graphene domains.[32] This growth mode has been reported in Ref.[18] for graphene growth on Ge(100) at atmospheric pressure. However, we do not observe the minimum in the nucleation probability of the second layer they found at intermediate CH$_4$ flow: in our samples the higher is the flow the higher is the multilayer nucleation and the smaller the domain size. The difference can reside in the more accurate control of temperature in our growths. While we fixed T=930 °C, in Ref.[18] the growth temperature was between 910 and 930 °C, a range in which temperature variations have a strong impact in the graphene synthesis.

CONCLUSIONS

We investigated the CVD growth of graphene on Ge(100) as a function of the methane flow, H$_2$:CH$_4$ flow ratio and deposition time. The combined analysis of the Raman, XPS and SEM/AFM data allowed us to tune the growth in order to obtain graphene nanoribbons, a layer by layer growth and multilayer graphene synthesis by simply varying the CH$_4$ flow. A layer-by-layer growth regime occurred for F=2 sccm and R=100. Employing these "special" growth conditions we obtained the reliable and well controlled growth of graphene monolayer with average I$_{2D}$/I$_G$ ratio larger than 3, a value exceeding the state of the art for graphene synthesis on Ge.

AUTHOR INFORMATION

**Corresponding Author**

\* prof. Monica De Seta

email: monica.deseta@uniroma3.it

Dipartimento di Scienze, Università Roma Tre, V. Guglielmo Marconi 446

00146 Rome Italy

**Author Contributions**

⊥M. D. S. and L. D. G. contributed equally to this work.

**Funding Sources**

The work was partially supported by the European Union Seventh Framework Program under grant agreement n° 604391 Graphene Flagship

**Notes**

The authors declare no competing financial interests

ACKNOWLEDGMENTS

The authors thank prof. A. Ruocco for the helpful discussions. The SEM measurements were performed at the Interdepartmental Laboratory of Electron Microscopy (LIME) of Roma TRE University.

**Table Of Contents (TOC):**

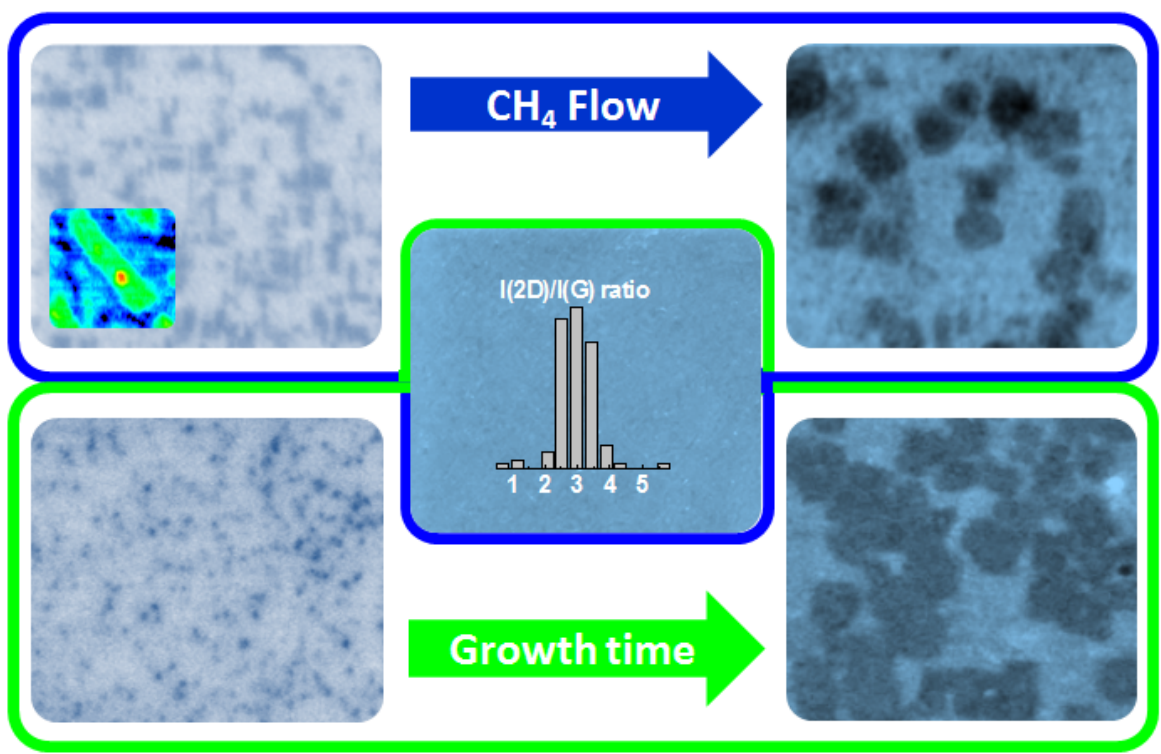